\documentclass[
10pt, 
a4paper, 
oneside, 
headinclude,footinclude, 
BCOR5mm, 
]{scrartcl}

\usepackage[
nochapters, 
beramono, 
eulermath,
pdfspacing, 
dottedtoc 
]{classicthesis} 

\usepackage{arsclassica} 

\usepackage[T1]{fontenc} 

\usepackage[utf8]{inputenc} 

\usepackage{graphicx} 
\graphicspath{{Figures/}} 

\usepackage{enumitem} 

\usepackage{lipsum} 

\usepackage{subfig} 

\usepackage{amsmath,amssymb,amsthm} 

\usepackage{varioref} 

\usepackage{geometry}

\usepackage{titlesec}

\titlespacing*{\section}
{0pt}{3.5ex plus 1ex minus .2ex}{2.3ex plus .2ex}

\usepackage{amsmath}

\usepackage{natbib}

\theoremstyle{definition} 

\theoremstyle{plain} 

\theoremstyle{remark} 

\hypersetup{
colorlinks=true, breaklinks=true, bookmarks=true,bookmarksnumbered,
urlcolor=webbrown, linkcolor=RoyalBlue, citecolor=webgreen, 
pdftitle={}, 
pdfauthor={\textcopyright}, 
pdfsubject={}, 
pdfkeywords={}, 
pdfcreator={pdfLaTeX}, 
pdfproducer={LaTeX with hyperref and ClassicThesis} 
} 

\title{\vspace{-1.5cm}\normalfont\spacedallcaps{Spectres: A fast spectral \protect \\resampling tool in Python}} 

\author{\spacedlowsmallcaps{A. C. Carnall} - adamc@roe.ac.uk} 

\date{\small{Institute for Astronomy, University of Edinburgh, Royal Observatory, Edinburgh EH9 3HJ}} 

\begin{document}


\renewcommand{\sectionmark}[1]{\markright{\spacedlowsmallcaps{#1}}} 
\lehead{\mbox{\llap{\small\thepage\kern1em\color{halfgray} \vline}\color{halfgray}\hspace{0.5em}\rightmark\hfil}} 

\pagestyle{scrheadings} 


\maketitle 

\setcounter{tocdepth}{2} 





\section*{Abstract} 

I present a fast Python tool, SpectRes, for carrying out the resampling of spectral flux densities and their associated uncertainties onto different wavelength grids. The function works with any grid of wavelength values, including non-uniform sampling, and preserves the integrated flux. This may be of use for binning data to increase the signal to noise ratio, obtaining synthetic photometry, or resampling model spectra to match the sampling of observed data for spectral energy distribution fitting. The function can be downloaded from \url{https://www.github.com/ACCarnall/SpectRes}.






\section{The SpectRes Code}

SpectRes is a Python function which efficiently resamples spectra onto an arbitrary wavelength grid using the method described in Sections \ref{problem}, \ref{sec2} and \ref{sec3}. To illustrate the speed of the code, a standard ised file from \cite{Bruzual2003} containing 221 models each with 6917 wavelength points is resampled onto a uniform 5\AA\ grid between 1000\AA\ and 11000\AA, as might be typical for spectral fitting, in a time of $\sim$ 40ms. An example result is shown in Figure \ref{fig2}. Resampling of a single model onto a uniform 20\AA\ grid between 3000\AA\ and 50000\AA, as might be typical for photometric fitting, takes $\sim$ 10ms. This makes the code highly applicable to, for example, rapid calculation of photometric redshifts. Figure \ref{fig3} shows the rebinning of a high redshift quasar spectrum from \cite{Carnall2015} in order to improve the signal to noise ratio.

The code is available for download from \url{https://www.github.com/ACCarnall/SpectRes}. The function takes three arguments, firstly \textbf{spec\_wavs}, an array of wavelengths corresponding to the current sampling of the spectrum. Secondly \textbf{spec\_fluxes}, a 1D or 2D array of flux values with the first axis running over wavelength and the second (if present) over different spectra to be resampled. The third argument, \textbf{resampling}, is the desired sampling. A keyword argument, \textbf{spec\_errs} may also be passed of the same shape as \textbf{spec\_fluxes} containing the uncertainty on each flux value. The function returns an array, \textbf{resampled} with first dimension the same length as \textbf{resampling} and second dimension (if present) the same length as the second dimension of \textbf{spec\_fluxes}. If the keyword argument \textbf{spec\_errs} is passed, a second array, \textbf{resampled\_errs} of the same shape, containing the resampled error spectra is also returned.

\section{Problem} \label{problem}

One dimensional astronomical spectra normally take the format of a set of wavelength values $\lambda_i$ ($i\ \epsilon$ 1, 2, ..., n) with associated flux per unit wavelength values $f_{\lambda i}$, or alternatively frequency values $\nu_i$ with associated flux per unit frequency values $f_{\nu i}$, along with uncertainties $\sigma_i$. For the remainder of this article we will assume a grid of wavelength values and associated flux per unit wavelength values, however the function presented should be equally applicable to either case.

\begin{figure}[t]
\centering 
\includegraphics[width=\columnwidth]{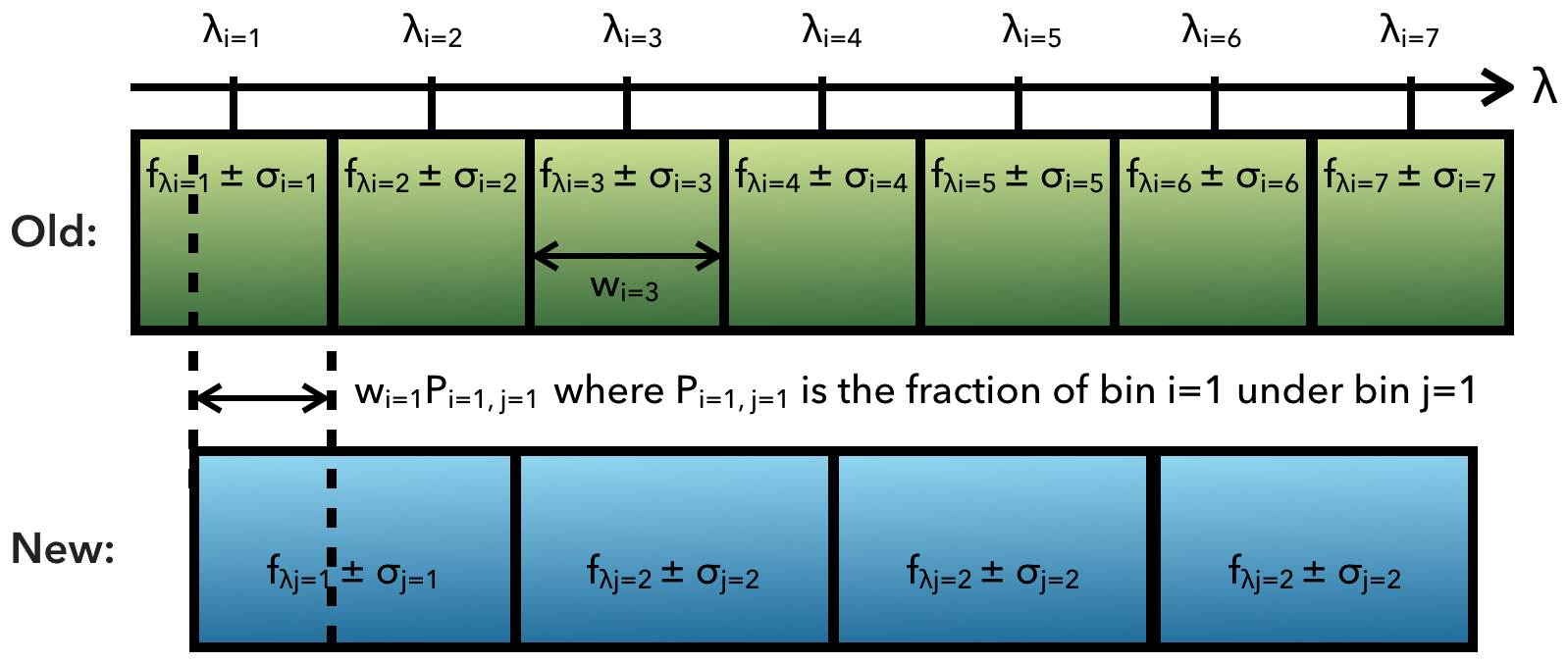} 
\caption{The configuration of old and new spectral bins as described in Section \ref{problem}}
\label{fig1}
\end{figure}

\begin{figure}[t]
\centering 
\includegraphics[width=\columnwidth]{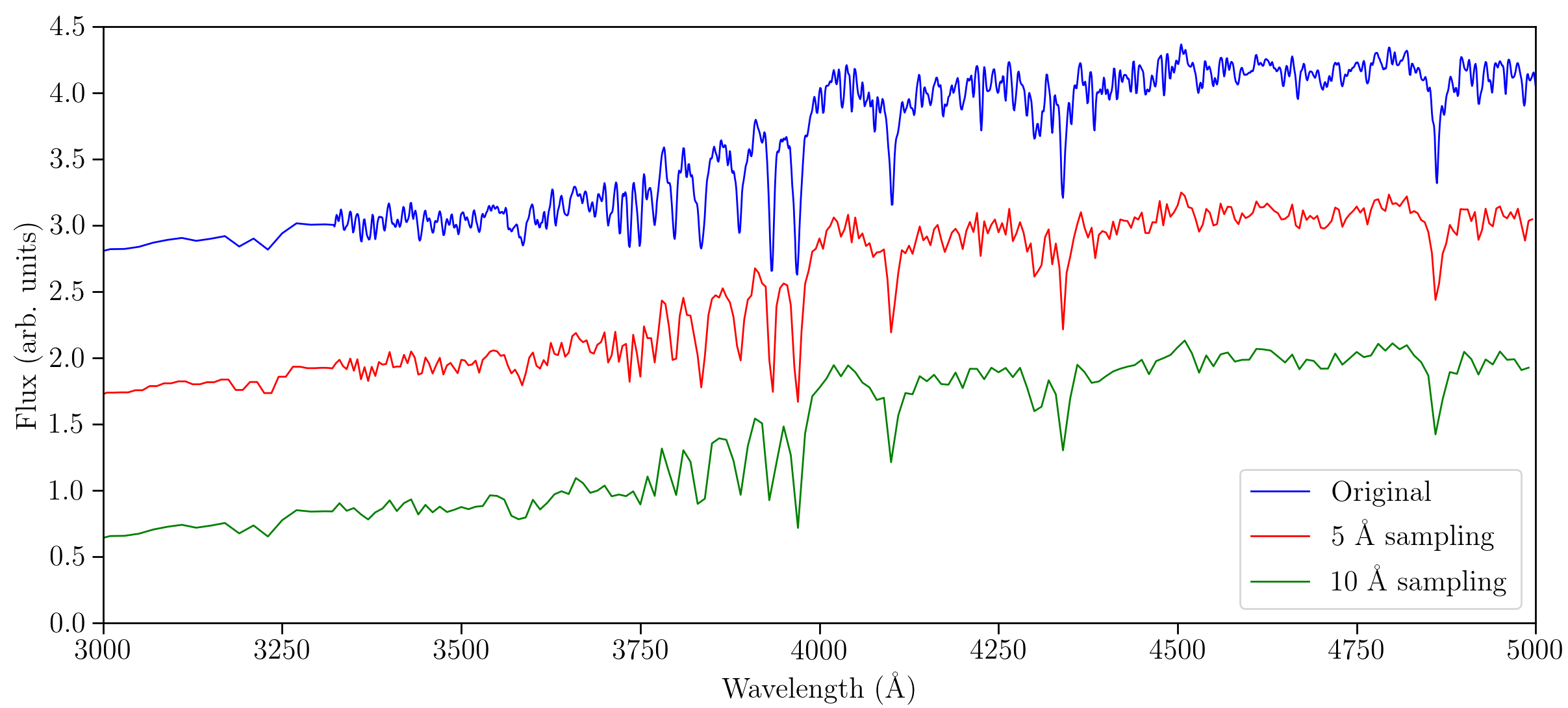} 
\caption{Spectral resampling of a model of a 1 Gyr old burst of star formation from \protect \cite{Bruzual2003} with SpectRes. The original model is shown in blue, the red model has been resampled onto a uniform 5\AA\ grid and the green model onto a uniform 10\AA\ grid. Vertical offsets have been applied for clarity.}
\label{fig2}
\end{figure}

\begin{figure}[t]
\centering 
\includegraphics[width=0.95\columnwidth]{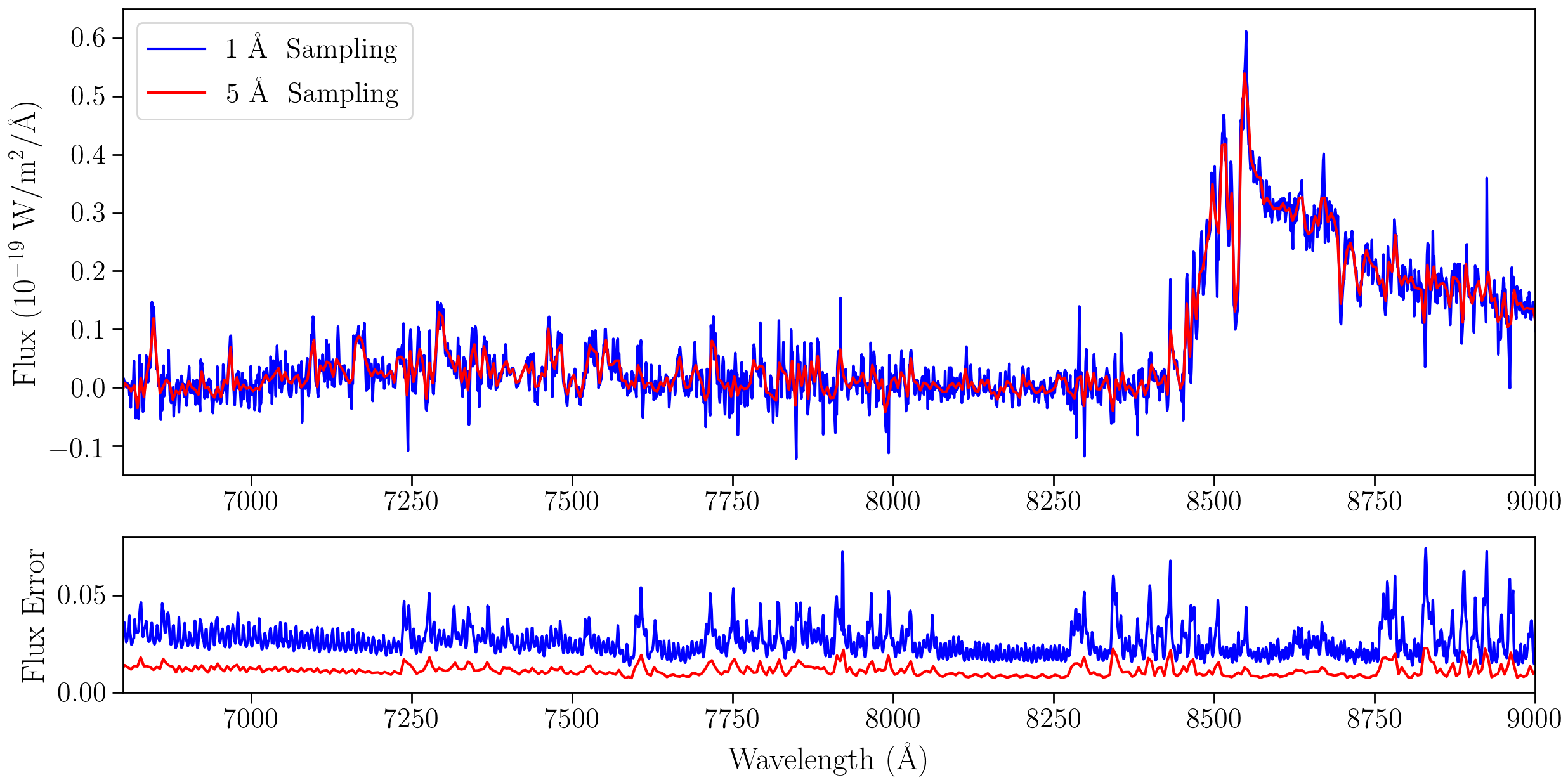} 
\caption{Spectrum of a high redshift quasar from \cite{Carnall2015} with original sampling at 1\AA\ in blue and resampled to 5\AA\ with SpectRes in red. The error spectra are shown below.}
\label{fig3}
\end{figure}

The units of $f_{\lambda i}$ are normally of the form energy per unit time per unit of collecting area per unit wavelength (e.g. erg/s/cm$^2$/\AA). Each $f_{\lambda i}$ value can be thought of as the average flux per unit wavelength value across the interval

\begin{equation}
\Big( \frac{\lambda_{i-1} + \lambda_{i}}{2}\  ,\  \frac{\lambda_{i} + \lambda_{i+1}}{2} \Big)
\end{equation}
\\
\noindent We therefore define $w_i$ as the width of each of these intervals in wavelength space. The subject of this article is to define another set of values $f_{\lambda j}$ and uncertainties $\sigma_j$ with different associated wavelength values $\lambda_j$ ($j\ \epsilon$ 1, 2, ..., m). The described configuration is shown graphically in Figure \ref{fig1}.

\section{Calculation of Flux Values} \label{sec2}

The first step in the calculation of the new bin flux values $f_{\lambda j}$ is to obtain a matrix of $P_{ij}$ values, which are the fraction of the wavelength range covered by each initial bin $i$ which also falls under the final bin $j$. A simple, yet computationally intensive method for doing this would be to calculate

\begin{align} \label{eqn2}
\begin{split}
(L_{i+1} - L_i)\ P_{ij}\ \  &=\ \ \  (L_{i+1} - L_{j})\ \ \ \ \ \ \ \ \ \  \mathrm{if}\ \ \  L_{j} > L_{i}\ \  \mathrm{and}\ \  L_{j+1} > L_{i+1} \\
           				    &=\ \ \  (L_{j+1} - L_{i})\ \ \ \ \ \ \ \ \ \  \mathrm{if}\ \ \  L_{j} < L_{i}\ \  \mathrm{and}\ \  L_{j+1} < L_{i+1} \\
           				    &=\ \ \  (L_{j+1} - L_{j})\ \ \ \ \ \ \ \ \ \  \mathrm{if}\ \ \  L_{j} > L_{i}\ \  \mathrm{and}\ \  L_{j+1} < L_{i+1} \\
           				    &=\ \ \  (L_{j+1} - L_{i})\ \ \ \ \ \ \ \ \ \  \mathrm{if}\ \ \  L_{j} < L_{i}\ \  \mathrm{and}\ \  L_{j+1} > L_{i+1}
\end{split}
\end{align}
\\

\noindent where $L_{i,j} = (\lambda_{i-1,j-1} + \lambda_{i,j})/2$ is the left hand edge of each bin, and any $P_{ij}$ value of less than zero should subsequently be set to zero. The method implemented in SpectRes is much more highly optimised, but more convoluted, so a discussion is not presented here. Interested parties should refer to the code.

Once the $P_{ij}$ values have been obtained, the values of the $f_{\lambda j}$ can be obtained by performing the weighted sum

\begin{equation} \label{eqn3}
f_{\lambda j}\ \ \  =\ \ \ \frac{\sum_{i=1}^{n} (P_{ij} w_i f_{\lambda i})}{\sum_{i=1}^{n} (P_{ij} w_i )}\ \ \ =\ \ \ \sum_{i=1}^{n} c_{ij} f_{\lambda i}.
\end{equation}

\section{Calculation of Associated Uncertainties} \label{sec3}

The variances on the $f_{\lambda j}$ values calculated using Equation \ref{eqn3} can be obtained using standard uncertainty propagation formulae. The variance $\sigma_j^2$ on each $f_{\lambda j}$ is given, assuming all $f_{\lambda i}$ are independent, by

\begin{equation} \label{eqn4}
\sigma_j^2\ \ \ =\ \ \ \frac{\sum_{i=1}^{n} P_{ij}^2 w_i^2 \sigma_{i}^2}{\big(\sum_{i=1}^{n} P_{ij} w_i \big)^2}\ \ \ =\ \ \ \sum_{i=1}^{n} c_{ij}^2 \sigma_{i}^2.
\end{equation}
\\
\noindent It is important to recognise that in any case where adjacent flux values $f_{\lambda j}$ have contributions from the same $f_{\lambda i}$, those flux values will be covariant with each other to some extent. To see this, consider the case in which one starts with a uniformly spaced set of $\lambda_i$, and a uniform error spectrum $\sigma_i = \sigma$, and offsets each bin by half of its width so that $\lambda_j = (\lambda_{i+1} + \lambda_{i})/2$. After resampling we now have the same number of points in our spectrum, but each uncertainty $\sigma_j = \sigma / \sqrt{2}$. Clearly the new flux values must be covariant, otherwise it would be possible to continue to reduce the errors by this process without impacting on the data quality. We now derive the covariance matrix for the new bins.

\section{The Covariance Matrix}

The covariance matrix is constructed of elements $\sigma_{jb}$, between the new flux values $f_{\lambda j} = \sum_{i=1}^{n} c_{ij} f_{\lambda i}$ and $f_{\lambda b} = \sum_{a=1}^{n} c_{ab} f_{\lambda a}.$ where ($j, b\ \epsilon$ 1, 2, ..., m) and ($i, a\ \epsilon$ 1, 2, ..., n).

To obtain the $\sigma_{jb}$ values we first express the flux values in the old and new bins as Gaussian random variables X and Y respectively, e.g. $Y_j \thicksim \mathrm{N}(f_{\lambda j}, \sigma_j^2)$. We can express the random variables representing the new bins in terms of the old bins using the standard result from probability theory that

\begin{equation}\label{eqn5}
Y_j\ \ \ =\ \ \ \sum_{i=1}^{n} c_{ij} X_i\ \ \thicksim\ \ \ \mathrm{N}(f_{\lambda j}, \sigma_j^2)\ \ \ =\ \ \ \mathrm{N}\Bigg(\sum_{i=1}^{n} c_{ij} f_{\lambda i}, \sum_{i=1}^{n} c_{ij}^2 \sigma_i^2\Bigg),
\end{equation}
\\
\noindent which can be seen to be in agreement with our Equations \ref{eqn3} and \ref{eqn4}. Now the covariance between the two random variables, $Y_j$ and $Y_b$ is defined as

\begin{equation}\label{eqn6}
\mathrm{cov}(Y_j, Y_b)\ \ \ =\ \ \ \sigma_{jb}^2\ \ \ =\ \ \ \mathrm{E}\Big[\big(Y_j - E[Y_j]\big)\big(Y_b - E[Y_b]\big)\Big]
\end{equation}
\\
\noindent where $\mathrm{E}[X]$ is the expectation value of $X$. This can be expressed using Equation \ref{eqn5} as

\begin{equation}\label{eqn7}
\sigma_{jb}^2\ \ \ =\ \ \ \mathrm{E}\Bigg[\bigg(\sum_{i=1}^{n} c_{ij} X_i - \mathrm{E}\Big[\sum_{i=1}^{n} c_{ij} f_{\lambda i}\Big]\bigg)\bigg(\sum_{a=1}^{n} c_{ab} X_a - \mathrm{E}\Big[\sum_{a=1}^{n} c_{ab} f_{\lambda a}\Big]\bigg)\Bigg].
\end{equation}
\\
\noindent The distribution of $(X - \mathrm{E}[X_i])$ is a Gaussian with mean of zero and variance of $\sigma_i^2$, i.e. $\mathrm{N}\big(0, \sigma_i^2\big)$, meaning that Equation \ref{eqn7} can be written as 

\begin{equation}\label{eqn8}
\sigma_{jb}^2\ \ \ =\ \ \ \mathrm{E}\Bigg[\bigg(\sum_{i=1}^{n} c_{ij}\ \mathrm{N}\big(0, \sigma_i^2\big)\bigg)\bigg(\sum_{a=1}^{n} c_{ab}\ \mathrm{N}\big(0, \sigma_a^2\big)\bigg)\Bigg].
\end{equation}
\\
\noindent Multiplying out the brackets and taking the expectation value will yield zero for all $i \neq a$, as $\mathrm{E}\big[\mathrm{N}\big(0, \sigma^2\big)\ \mathrm{N}\big(0, \sigma^2\big)\big] = 0$. We can therefore simplify Equation \ref{eqn8} to

\begin{equation}\label{eqn9}
\sigma_{jb}^2\ \ \ =\ \ \ \mathrm{E}\Bigg[\sum_{i=1}^{n} c_{ij}\ c_{ib}\ \mathrm{N}\big(0, \sigma_i^2\big)^2\Bigg]\ \ \ =\ \ \ \sum_{i=1}^{n} c_{ij}\ c_{ib}\ \mathrm{E}\Big[\mathrm{N}\big(0, \sigma_i^2\big)^2\Big].
\end{equation}
\\
\noindent Finally we can use another standard result, that the variance, Var$[X] = \mathrm{E}[X^2] - \mathrm{E}[X]^2$ to obtain

\begin{equation}\label{eqn10}
\sigma_{jb}^2\ \ \ =\ \ \ \sum_{i=1}^{n} c_{ij}\ c_{ib}\ \sigma_i^2,
\end{equation}
\\
\noindent which agrees with the result of Equation \ref{eqn4} in the case where $j = b$ as expected. In this way the covariance matrix can be calculated, however where possible it is advisable to resample the model rather than observed spectrum in order to avoid dealing with covariances when performing spectral energy distribution fitting.

\section*{Acknowledgements} 

I would like to thank Raphael Errani, David Homan, Joseph Kennedy, Fran Lane, Andy Lawrence, Ross McLure and Tom Shanks for helpful discussions. The LaTeX template used was downloaded from http://www.LaTeXTemplates.com.

\renewcommand{\refname}{\spacedlowsmallcaps{References}} 

\bibliographystyle{mnras} 
\setlength{\bibsep}{0pt plus 0.3ex}
\bibliography{resampling} 


\end{document}